**Pseudoelastic deformation in Mo-based refractory multi-principal element alloys**

A. Sharma,[1,2,*] P. Singh,[1,*] T. Kirk,[3] V.I. Levitas,[4,5,6] P.K. Liaw,[7] G. Balasubramanian,[8] R. Arroyave,[3,9] and Duane D. Johnson[1,6]

[1]*Ames Laboratory, U.S. Department of Energy, Iowa State University, Ames, IA 50011 USA*
[2]*Sandvik Coromant R&D, Stockholm, 12679 Sweden*
[3]*Department of Materials Science & Engineering, Texas A&M University, College Station, TX 77843, USA*
[4]*Department of Mechanical Engineering, Iowa State University, Ames, IA 50011 USA*
[5]*Department of Aerospace Engineering, Iowa State University, Ames, IA 50011 USA*
[6]*Department of Materials Science & Engineering, Iowa State University, Ames, IA 50011 USA*
[7]*Department of Materials Science & Engineering, The University of Tennessee, Knoxville, TN, 37996 USA*
[8]*Department of Mechanical Engineering & Mechanics, Lehigh University, Bethlehem, PA 18015 USA*
[9]*Department of Mechanical Engineering, Texas A&M University, College Station, TX 77843, USA*

## Abstract

Phase diagrams supported by density functional theory methods can be crucial for designing high-entropy alloys that are subset of multi-principal-element alloys. We present phase and property analysis of quinary $(MoW)_xZr_y(TaTi)_{1-x-y}$ refractory high-entropy alloys from combined Calculation of Phase Diagram (CALPHAD) and density-functional theory results, supplemented by molecular dynamics simulations. Both CALPHAD and density-functional theory analysis of phase stability indicates a Mo-W-rich region of this quinary has a stable single-phase body-centered-cubic structure. We report first quinary composition from Mo-W-Ta-Ti-Zr family of alloy with pseudo-elastic behavior, i.e., hysteresis in stress-strain. Our analysis shows that only Mo-W-rich compositions of Mo-W-Ta-Ti-Zr, i.e., Mo+W$\geq 85\ at.\%$, show reproducible hysteresis in stress-strain responsible for pseudo-elastic behavior. The $(MoW)_{85}Zr_{7.5}(TaTi)_{7.5}$ was down-selected based on temperature-dependent phase diagram analysis and molecular dynamics simulations predicted elastic behavior that reveals twinning-assisted pseudoelastic behavior. While mostly unexplored in body-centered-cubic crystals, twinning is a fundamental deformation mechanism that competes against dislocation slip in crystalline solids. This alloy shows identical cyclic deformation characteristics during uniaxial <100> loading, i.e., the pseudoelasticity is isotropic in loading direction. Additionally, a temperature increase from 77 to 1,500 K enhances the elastic strain recovery in load-unload cycles, offering possibly control to tune the pseudoelastic behavior.

---

*AS and PS equally contributed

## Introduction

High-entropy alloys (HEAs), or more general multi-principal-element alloys (MPEAs), have attracted immense interest for their vast unexplored design space with the possibility of exceptional properties [1-13]. The formation of a stable, single-phase solid solution (face-centered-cubic (fcc), body-centered-cubic (bcc), or hexagonal-close-packed (hcp)) in a compositionally complex system with four or more elements at 5-35 *at.%* is an attractive and sometimes desirable characteristic of MPEAs, with stability often claimed to be due to entropic effects [1]. In practice, however, phase stability is the result of competition at the Gibbs level, and the configurational entropy alone is not always a dominant contributor towards phase stability [14]. The development of MPEAs has helped resolve some shortcomings of conventional alloys [7,15], but also raised fundamental questions that remain unexplained [16-22]. The high entropy is the main concept behind the formation solid-solution alloys [1]. However, recent works show that it is critical not to limit design space unnecessarily [13,23].

Recently, refractory MPEAs (RMPEAs) have gained considerable interest due to their high-temperature strength, resistance to fatigue, (sometimes) oxidation, corrosion, and wear [23-31]. However, there have been no reports on pseudoelastic behavior in RMPEAs, as they are predominantly seen in shape-memory alloys (SMAs) [32,33], like Ni-Ti-based alloys. A pseudoelastic material finds potential uses in electromechanical devices, including energy harvesting, defense equipment, and naval architecture [28-31]. Unlike NiTi-based SMAs, refractory-based elements, e.g., Mo, W, Ta, Ti, and Zr, are non-toxic, offering a reason to look beyond standard SMAs [34,35]. Pseudoelasticity in SMAs is the result of first-order martensitic (diffusionless) transformations [36], consisting of complex microstructures, including twinned martensites, wedges, and twins within twins [37,38]. Although, twinning-induced psuedoelasticity has been reported in refractory-based nanostructured elemental tungsten [39], no such reports are found in RMPEAs.

Here, using density-functional theory (DFT), molecular dynamic (MD), and CALPHAD calculations, we provide a stability, phase-diagram, and property analysis of $(MoW)_xZr_y(TaTi)_{1-x-y}$ composed of non-toxic elements from the periodic table from groups IV-A (Ti, Zr), V-A (Ta), and VI-A (Mo, W). The down-selection of Mo-W-rich alloys, specifically $(MoW)_{85}Zr_{7.5}(TaTi)_{7.5}$ (denoted here as MWZTT), was done based on thermodynamic phase stability analysis from CALPHAD and DFT that shows stable bcc in Mo-W-rich region (away from the near-equiatomic region of the 5-element system); on the other hand, W-rich alloys were found with interesting properties, such as shape-memory [39] and higher-radiation resistance [40]. Hence, we did investigate the entire MPEA system, where high-entropy regions are much less stable, and other regions have critical electronic stabilizing effects [23]. As the configurational entropy of MWZTT is $S_{conf}$ = 1.17R, at the boundary of medium-entropy alloys (MEAs) and HEAs (1.5R > $S_{conf}$(MEAs) > 1R), we categorized MWZTT as the part of high-entropy alloy family. The MWZTT shows recoverable non-linear elastic (pseudo-elastic) behavior under uniaxial loading, as obtained by MD and supported by our finite-element analysis. The MD-determined phase and mechanical stability are in good agreement with our DFT results, while CALPHAD shows a stability of bcc phase up to 1,700 K. We also show that MWZTT is structurally stable, i.e., dynamical stable having positive phonon frequencies. The present work opens unexplored applications of RMPEAs using unique pseudoelastic properties found in small regions of composition space.

## Computational Method

### Density-Functional Theory (DFT) methods

(i) *<u>Electronic-structure methods:</u>* Two methods were employed, i.e., an *all-electron KKR-CPA* (a multiple-scattering theory Green's function) method that directly averages over chemical configurations to all orders in an infinite crystal concomitantly with DFT charge self-consistency permitting primitive cells to be utilized, and *a plane-wave pseudo-potential method* (VASP) using a representative supercell (described in (ii)). KKR electronic-structure energy calculations utilize an atomic sphere approximation [41], where chemical disorder in MPEAs was handled using the coherent-potential approximation (CPA) [42,43]. The KKR-CPA method is an effective-medium theory that solves for an effective (complex) scatterer at a site to recover the scattering properties on average for the collective disordered system. The effects of disorder in DFT-based KKR-CPA are represented by broadening the electronic structure [41-43], and not possible in an effective-atom method (see supplemental material). A gradient-corrected exchange-correlation functional (PBE) were employed in all calculations [44]. At each atomic-sphere site, periodic boundary conditions along with a Voronoi polyhedral scheme was used for spatial integrals to account for an accurate unit cell charge distribution and energy calculation, yielding energy difference similar to full-potential all-electron methods for ordered alloys as well as partially ordered and fully disordered alloys [45,46]. The Green's function was solved on a semi-circular complex-energy contour that uses a Gauss-Laguerre quadrature (with 24 complex energies enclosing the bottom to the top of the valence states). A Monkhorst-Pack [47] $\boldsymbol{k}$-point mesh of $20 \times 20 \times 20$ was used for Brillouin zone (BZ) integrations for self-consistent calculations, and, for care, $50 \times 50 \times 50$ $\boldsymbol{k}$-point mesh was employed for the total density of states (DOS). Bloch-spectral-functions (BSF) were calculated for 300 k-points to visualize the electronic dispersion along high-symmetry lines of bcc BZ. The BSF is defined in terms of energy (E) and reciprocal-space (k) resolved dispersion, which reduces to Dirac delta function $\delta(\epsilon\text{-}E_k)$ in the limit of an ordered system [48]. The CPA inherently includes spectral broadening and shifting due to chemical disordered effects and composition.

(ii) *Mechanical and structural stability determination*: The elastic properties of MWZTT were calculated in a DFT-based stress-tensor approach as implemented in the plane-wave pseudo-potential Vienna *Ab-initio* Simulation Package (VASP) [49,50]. The Perdew, Burke, and Ernzerhof (PBE) generalized gradient approximation [44] to DFT was used with a energy cut-off of 533 eV. For the supercells, a Monkhorst-Pack [47] k-mesh (2 x 2 x 4) was employed for Brillouin zone integration during structural-optimization and charge self-consistency. The energy and forces were converged, respectively, to $10^{-6}$ eV and $-10^{-6}$ eV/ Å. Six finite distortions were performed for stress-strain calculations to derive the elastic moduli ($C_{ij}$) [51,52]. An 80-atom Super-Cell Random Approximates (SCRAPs) was used to mimic the homogeneously random alloy (here, using bcc 2-atom cell with periodicity of 5 x 4 x 2). The SCRAP was generated using a Hybrid Cuckoo Search optimization method/code [53] for ultrafast creation for arbitrary MPEAs and any size supercell with specified point and pair correlations, i.e., elemental compositions and pairwise atomic short-range order, respectively. In the assumed SCRAPs, phonon dispersion to determine structural stability was calculated using density-functional perturbation theory (DFPT), as implemented in VASP [54-56]. Dispersion was plotted along high-symmetry points of a bcc BZ (G-N-P-G-H).

**CALPHAD**: Isothermal equilibrium phase diagrams were calculated via CALPHAD (CALculation of PHAse Diagrams) techniques [57-60], using Thermo-Calc's TCHEA4 High-Entropy Alloy Database [61]. In this database, 6 of the 10 constituent ternaries of the Mo-W-Zr-Ta-Ti quinary system have been assessed (i.e., fit to the empirical data) in the full range of compositions and temperatures. One other ternary (TaWZr) has been assessed for a partial set of potential conditions. Three remaining ternaries (MoTaW, MoTaZr, and MoWZr) have not been assessed, but can be extrapolated from other data. These ternary assessments provide empirical support to the CALPHAD models and ensure that predictions are accurate, with possible exceptions in unassessed areas. Two-dimensional phase diagrams were created by constraining compositions to a function of two variables (x,y), i.e., $(MoW)_xZr_y(TaTi)_{1-x-y}$. Thermo-Calc's global minimization scheme was used to optimize phase Gibbs energies and identify stable phases in this constrained composition space at fixed temperatures.

**Molecular Dynamics (MD)**: We employ the Large-scale Atomic/Molecular Massively Parallel Simulator (LAMMPS) [62] software to perform controlled MD simulations [63]. Initially, a cuboidal simulation domain (**Fig. 1**) of 54,000 atoms was constructed by a random distribution of Mo, Ta, W, Ti, and Zr atoms in a bcc lattice of $(95.91 \times 95.91 \times 95.91)$ Å$^3$ using a 2-atom bcc cell of lattice constant of 3.19 Å [23]. In LAMMPS, we use specific "thermostats" to control the temperature (T) or pressure (P). We also performed common neighbor analysis (CNA) [64], using the Open Visualization Tool (OVITO) [65]. Periodic-boundary conditions were imposed in all three coordinate axes. The intermolecular interactions are described employing the assimilated Embedded Atomic Method (EAM) potentials [62,66] and validated with first-principles results [23] (see **Table 1**).

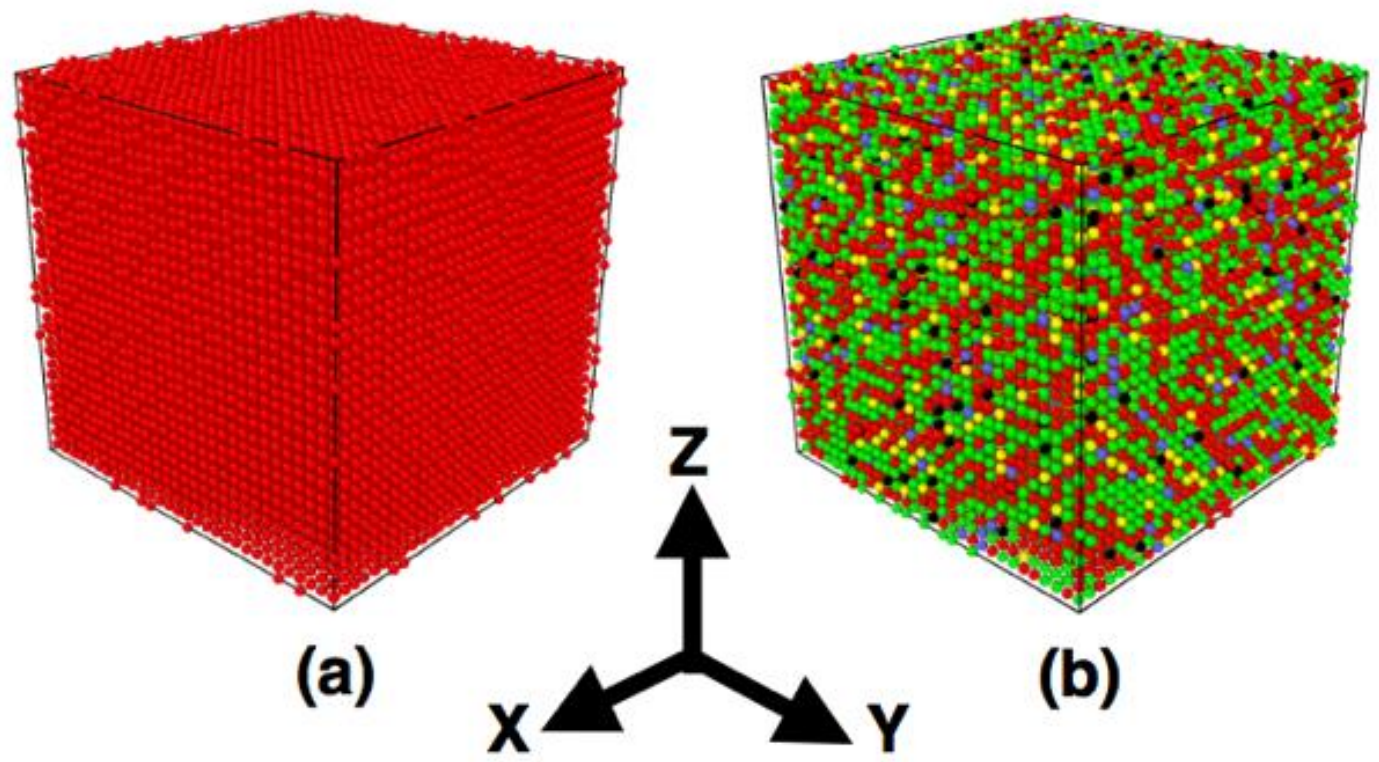

**Figure 1**. Simulation cells (54,000 atoms): (a) Tungsten (W) and (b) quinary MWZTT. Common neighbor analysis reveals bcc coordination in (a) and (b). Elements are visualized by colors (Mo = green, W = red, Ta = yellow, Ti = black, and Zr = blue).

The energy minimization of the structure was done using the conjugate-gradient algorithm with energy [force] tolerance of $10^{-6}$ eV [$-10^{-9}$ eV/Å] to find a geometrically optimized configuration. We quenched the RMPEAs from a high temperature (4,000 K) to lower temperatures (T=77, 300, 500, 700, 1100 K) using an isothermal-isobaric NPT process, which is followed by an NVE condition (microcanonical ensemble). This completes the equilibration of the MWZTT at a predefined temperature (additional details are provided in the supplement).

To understand the deformation mechanism in MD, we apply quasi-static uniaxial loading-unloading in the <100> direction. At each loading step, the simulation box is expanded in the <100> direction at a strain rate of 0.01 $ps^{-1}$, while the lateral boundaries are controlled to zero pressure using the NPT. The strain expressed on the sample is the true strain. Subsequently, the deformed alloy is equilibrated after each loading step under NPT and NVE ensembles for 90 and 50 ps, respectively. The MWZTT RMPEA is initialized at 4,000 K under an isothermal-isobaric (NPT) ensemble at a pressure of 0 MPa for 90 picoseconds (ps). Then the system is rapidly quenched from 4,000 K to different temperatures (300 to 1,500 K) in 10 ns, and then equilibrated for an additional 90 ps. The quenched RMPEAs are further simulated under the NPT and NVT (canonical) ensembles, successively for 10 and 20 ps, respectively. The pressure (0 MPa) and temperature constraints are imposed by the Nóse-Hoover barostat and thermostat with coupling times for both set at 1 ps. Finally, the structure is equilibrated for 10 ps under a microcanonical ensemble (NVE) to complete the quenching process. A time step of 0.001 ps is maintained throughout all simulations.

**Limitations of MD**: Finite length (nanometers, nm) and timescales (nanoseconds, ns or picoseconds, ps) are two important aspects in atomistic or MD simulations that need to be understood. Hence, higher strain rates are typically involved to study any deformation behavior using MD simulations. For example, at each step increase in strains, a limited number of energetic barriers can be overcome due to higher strain rates in MD. This happens due to finite time available for system to reach other configurations that are thermodynamically possible. To mitigate this, we have used quasi-static loading where each alloy relaxes further at each step of strain that could allow to reach thermodynamically stable configurations during both loading-unloading. The MD force-fields or atomistic potential are often the other constraint in modeling alloy systems. Benchmarking with available existing experiments or comparison against DFT are employed to test the feasibility of the potential. We have compared MD predictions of phase, structural and mechanical properties versus DFT, thermodynamic analysis (Calculation of Phase Diagrams, CALPHAD) and available experiments to provide confidence in our results.

## Results and discussion

A. *RMPEA Phase Stability*: Investigating the thermodynamic phase stability is the fundamental to understand alloy formation. However, the phase diagrams of RMPEAs can be time consuming if solely done using DFT and experiments. The CALPHAD-assessed temperature-dependent phase stability is done as a sweep through $(MoW)_x(TaTi)_yZr_{1-x-y}$ compositions and evaluated the phase-diagram at 300 K (room temperature), 700 K, 1,100K, and 1,700 K. In **Fig. 2**, the CALPHAD-assessed phases versus composition at room temperature can be divided mainly into four regions. However, only regions near Mo-W show bcc phase stability without the presence of hcp phases at lower temperatures (300 and 700 K). The RMPEAs in Mo-W-rich regions involve only one bcc phase with small amounts of Laves phase, which increase as other elements are added and are present in most of the composition space. As Ta, Ti and Zr are added, multiple bcc and hcp phases become stable. As we focused on a single-phase solid solution in the present work, the MoW-rich region was deemed most promising for design control. As 3 of the 10 constituent ternary systems of the MoWTaTiZr quinary have not been assessed in the CALPHAD database (TCHEA4), DFT formation energies were combined with CALPHAD-generated phase diagrams to enhance the accuracy of stability prediction.

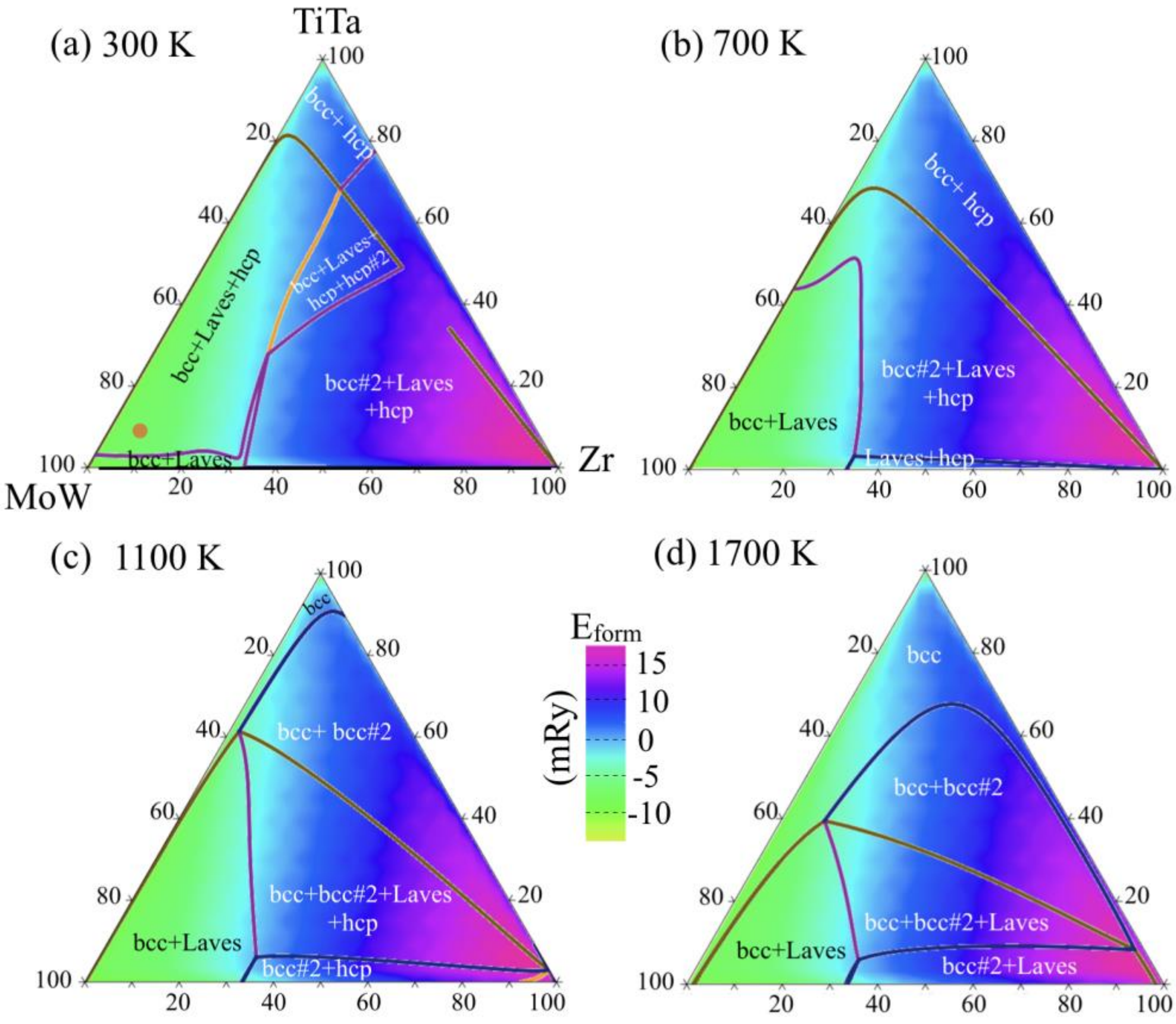

**Figure 2**. CALPHAD phase-diagram analysis and DFT formation energies (color bars) of $(MoW)_xZr_y(TaTi)_{1-x-y}$ RMPEAs. Here, (x,y) are given in atomic percent (at.%).

Based on a formation energy range ($-15 \leq \Delta E_f \leq +5$ mRy) calculated from DFT-KKR-CPA and the stability criterion involving valence-electron count ($4 < VEC < 6$) and atomic-size effect ($\leq$ 6.6%), we found that the MoW-rich region in **Fig. 2** (cyan and green region) of RMPEAs is energetically more stable in bcc than fcc or hcp [23]. Our CALPHAD phase diagram analysis (**Figs. 2a-d**) at 300-1,700 K also agrees with DFT predictions that shows that the bcc phase stability improves as we move to the Mo-W rich region in **Fig. 2**. Therefore, we focus mainly on the solid-solution phase of the phase diagram and choose an alloy composition $(MoW)_{85}(TaTi)_{7.5}Zr_{7.5}$ from the Mo-W rich region. Here, Mo-W gives high room-temperature strength and greater melting temperature. Moreover, Ti or Zr add strength at higher temperatures and provide lower density to the alloy, whereas Ta adds the electronic stability and high melting temperature.

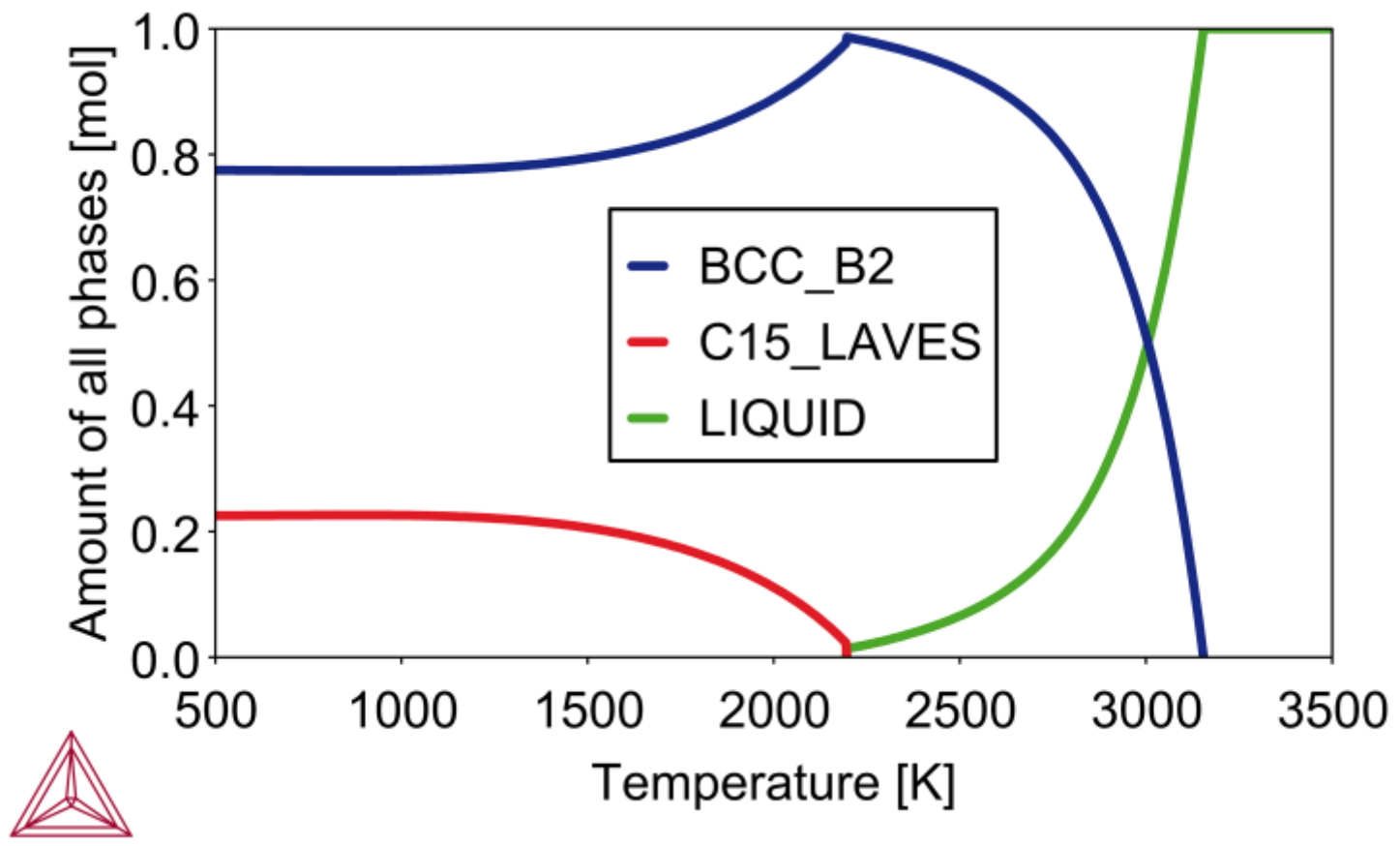

**Figure 3.** The equilibrium phases of the $(MoW)_{85}(TaTi)_{7.5}Zr_{7.5}$ RMPEA as a function of temperature.

The equilibrium phases versus temperature for MWZTT is presented in **Fig. 3**, as assessed by CALPHAD (TCHEA4). The solidus and liquidus temperatures are estimated to be 2200 K and 3160 K, respectively. At the solidus, the alloy is predicted to be a single-phase bcc. Nevertheless, as temperature decreases, the Laves phase is predicted to grow to a maximum of about 0.2 molar fraction. Formation of the Laves phase (rich in MoWZr) could be suppressed, however, if the solidification of the alloy is faster than the kinetics of the phase allow.

*B. Electronic-structure and structural properties of MWZTT*: The MD-calculated lattice constant (3.28 Å) fitted using Birch-Murnaghan equation of state (**Fig. 4a**) shows agree with DFT results (KKR-CPA: 3.165 Å and VASP-SCRAPs: 3.180 Å). The phonon dispersion (**Fig. 4b**) has only positive-definite modes (frequencies), indicating that the RMPEA is dynamically stable. The electronic dispersion and total DOS of MWZTT (**Figs. 4c-d**) show that the Fermi energy is in a pseudogap where the bonding (and non-bonding) states are filled, and anti-bonding states are empty. In **Fig. 4c**, sharp bands near -6.8 eV below the Fermi energy indicates that these states do not participate in hybridization in MWZTT. However, bands near the Fermi energy are much more diffuse (broadened in energy by compositional disorder), offering more scattering due to increased disorder, which will also affect resistivity. This trend also shows that only states near the Fermi energy are chemical active and hybridize with other elements. The minima in the total DOS at the Fermi energy is from *d*-electrons of constituent elements that lead to a pseudo-gap region [67,68], which is used to assess energy stability in MPEAs [31]. Phase stability (**Fig. 2**), phonons (**Fig. 4b)**, and electronic structure (**Fig. 4c-d)** analysis show that MWZTT should be of a single phase and stable.

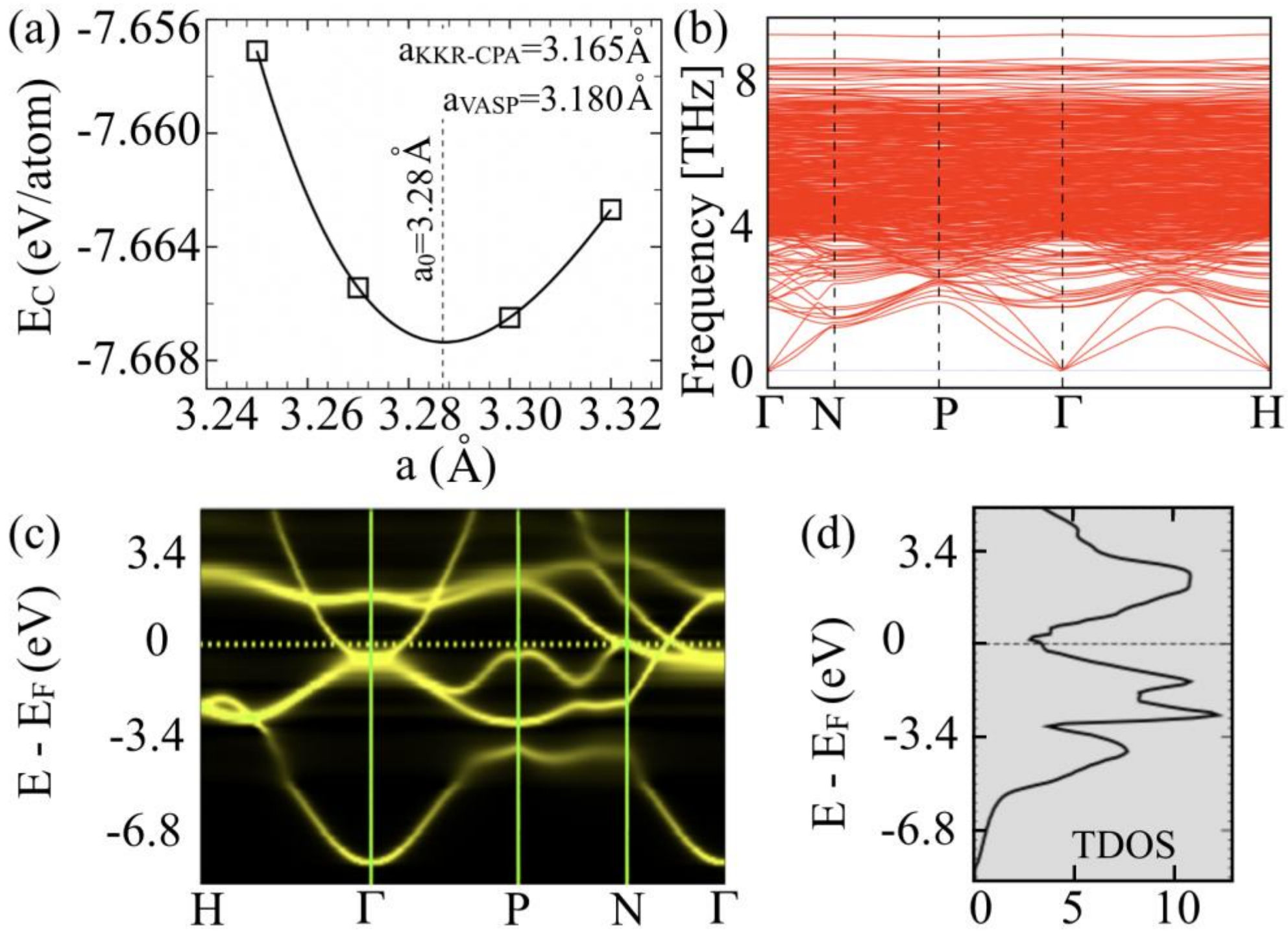

**Figure 4**. For MWZTT, (a) MD-calculated cohesive energy (meV/atom) vs. lattice constant (Å) was fitted to Birch-Murnaghan equation of state (at 77 K) to get equilibrium $a_o$=3.28 Å, in agreement with DFT results (KKR-CPA: 3.165 Å and VASP-SCRAP: 3.180 Å). (b) Using VASP-SCRAP, phonons on lattice with $a_o$ indicates it is dynamically stable, as all frequencies are positive. (c) Using KKR-CPA, electronic dispersion along high-symmetry lines of the bcc Brillouin zone, and the (d) total density of states (States-$eV^{-1}$-$atom^{-1}$). Similar TDOS (not shown) was found both from KKR-CPA (1-atom configurationally averaged) and VASP-SCRAP (80-atom, single configuration), as expected.

The elements of the elastic tensor ($C_{ij}$) can be used to determine elastic stability of materials, also permitting estimates of key physical properties, such as elastic moduli and Poisson's ratio, as well as Debye/melting temperatures. The standard criteria to determine elastic stability of cubic crystals involves

$$\text{(i) } C_{11} > 0, \text{ (ii) } C_{11}-|C_{12}| > 0, \text{ (iii) } C_{11}+2C_{12} > 0, \text{ and (iv) } C_{44} > 0 \qquad \text{Eq. (1)}$$

or in terms of bulk modulus ($B$), tetragonal shear modulus ($C'$) and shear modulus $C_{44}$ [69],

$$B = \frac{1}{3}(C_{11} + 2C_{12}) > 0; \; C' = \frac{1}{2}(C_{11} - C_{12}) > 0 \qquad \text{Eq. (2)}$$

Here, $C_{11}$, $C_{12}$, and $C_{44}$ are the three independent elastic constants for bcc MWZTT that relates six components of the stress (σ) tensor with those of the strain (ε) tensor. In **Table 1**, we provide the calculated lattice constant $a_o$, $B$, ($C_{11}$, $C_{12}$, $C_{44}$), and E (Young's moduli) for MWZTT.

**Table 1**: Calculated lattice constant ($a_o$ in Å**)**, cohesive energy ($E_c$ in eV), density (g/cm$^3$), and elastic constants (GPa) at 77, 300, 500, 700 and 1,100 K for MWZTT. Poisson's ratio (ν= 0.299) was directly calculated from DFT-VASP-SCRAPs for the modulus (E) calculation [70], and the bulk modulus B and density from KKR-CPA or VASP-SCRAPs agree, differing due to slightly different $a_o$ predicted.

| **T (K)** | ***$a_o$* (Å)** | **$\rho$ (g/cm$^3$)** | ***$E_c$* (*e*V)** | ***B* (GPa)** | **$C_{11}$ (GPa)** | **$C_{12}$ (GPa)** | **$C_{44}$ (GPa)** | **E (GPa)** |
|---|---|---|---|---|---|---|---|---|
| **DFT (VASP-SCRAPs)** | | | | | | | | |
| **0 K** | 3.180 | 13.73 | -- | 250 | 380 | 186 | 116 | 301 |
| **MD (LAMMPS)** | | | | | | | | |
| **77** | 3.28 | 13.68 | -7.66 | 251 | 395 | 179 | 123 | 303 |
| **300** | 3.25 | 13.76 | -- | 250 | 381 | 181 | 124 | 301 |
| **500** | 3.22 | 13.92 | -- | 248 | 367 | 184 | 125 | 299 |
| **700** | 3.19 | 14.35 | -- | 245 | 355 | 188 | 127 | 295 |
| **1100** | 3.14 | 14.63 | -- | 239 | 336 | 191 | 130 | 288 |
| **DFT (KKR-CPA)** | | | | | | | | |
| **0 K** | 3.165 | 13.85 | -- | 254 | -- | -- | -- | 306 |

The elastic moduli were calculated using the small-displacement method [[redacted]] with a finite deformation of 0.015 Å, which are applied to each atom independently along x,y, and z directions to extract elastic moduli. The results from both DFT and MD simulations are shown in **Table 1**. MWZTT satisfies the elastic stability criteria in Eq. 1 and 2. The elastic parameters in **Table 1** calculated from DFT (0K, VASP) and MD (77 K, LAMMPS) are in good agreement. The temperature dependence of $C_{ij}$ was also calculated from MD. A significant drop, from 395 GPa at 77 K to 336 GPa at 1100 K, was noted in $C_{11}$, while $C_{12}$ (179 → 191 GPa) and $C_{44}$ (123 → 130 GPa) was weakly increased. Physically, $C_{11}$ indicates a reduction in tensile elasticity, whereas $C_{12}$ and $C_{44}$ indicate a slow increase in lateral and shear characteristics, respectively. A longitudinal strain produces a change in volume without a change in shape, reflecting a larger change in $C_{11}$. In contrast, $C_{12}$ and $C_{44}$ are less sensitive to shearing than $C_{11}$. The tetragonal shear modulus drops due to a reduction in $C_{11}$ from 108 GPa at 77 K to 60.5 GPa at 1,100 K.

C. *Mechanical properties under tensile loading:*

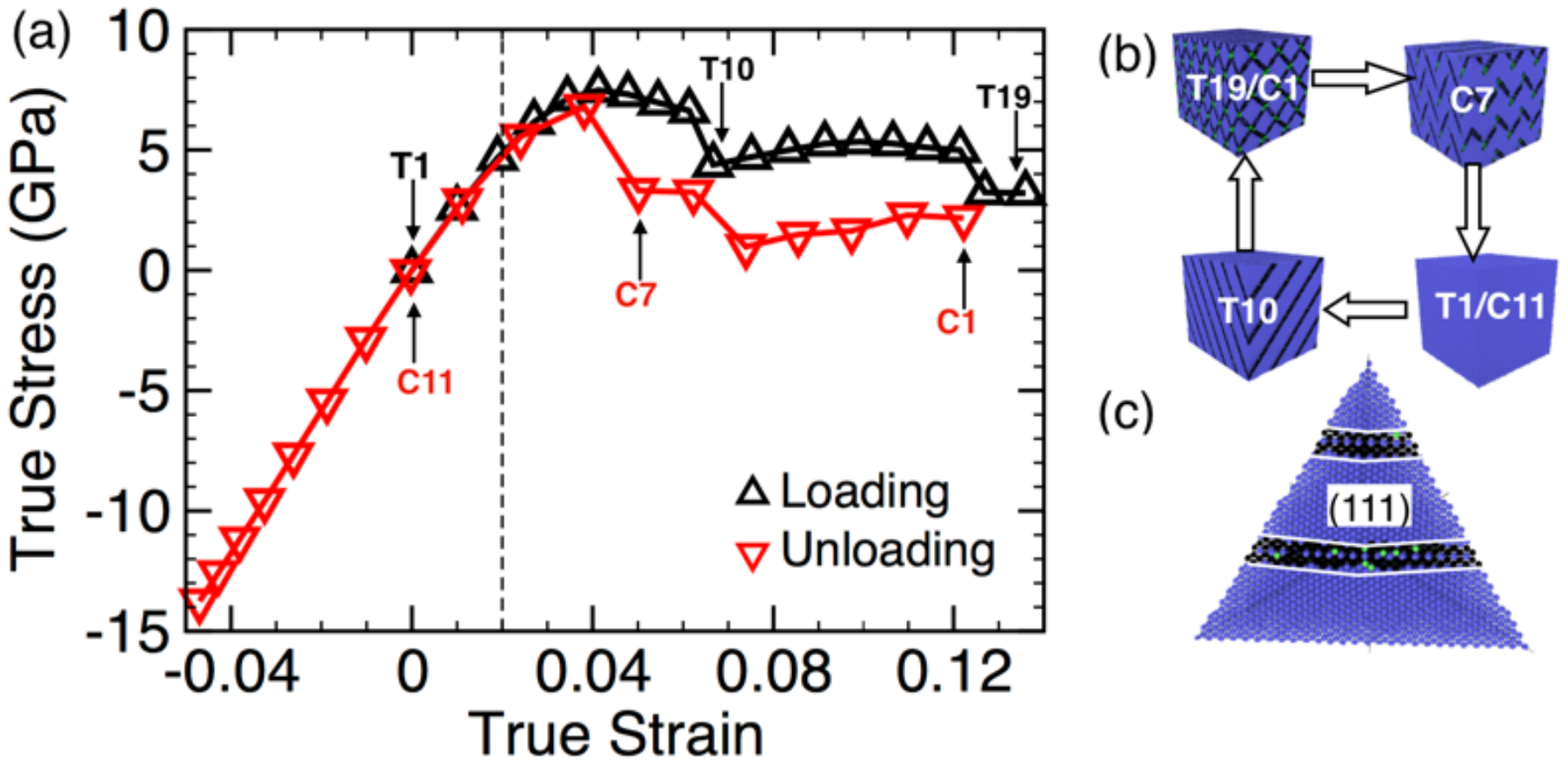

**Figure 5**. (a) Hysteresis found in quasistatic (un)loading cycle at 300 K in MWZTT. (b) Stepwise atomic-scale picture of twinning (loading) and detwinning (compressive loading). (c) For first twins, a (111) cut in the simulation cell shows the twin-plane with a twin-direction of <112>. In (a), twin formation occurs at a true strain of 0.065, and further strain-hardening yields cross-twins (T19). Compression (unloading), associated with a rise in the stress, leads to detwinning (C1 to C11).

To investigate the behavior under external strains, we performed a quasistatic tensile loading and compressive loading (unloading) on MWZTT. Uniaxial loading is widely used to reveal the deformation characteristics of a material. The cuboidal simulation domain showed a perfect BCC coordination after the structure had been quenched. Hence, there no imperfections such as dislocations or vacancies were found during the start of the loading process. One such deformation curve is presented in **Fig. 5** at 300 K. We show that the instability drives MWZTT RMPEAs away from the elastic regime above a strain of 0.02 and a stress of 6.16 GPa. As a stress piles up, twinning is observed in the crystalline lattice. The first major stress drop, where twinning relieves the stress, is observed in **Figs. 5a-b** (T10). This feature shows a deviation from a perfect bcc coordination. We observe a set of cross-twins in **Fig. 5**-T19 with increase in load, which disappears during unloading in **Figs. 5a-b** (C7). The CNA analysis confirms the twinning and tracks the nucleation of twins (their growth) and detwinning to a loading/unloading cycle. The MWZTT (**Figs. 5a-b** from C8-to-C11) follows the elastic regime during loading cycle, where detwinning is characterized by an abrupt rise in the stress level. The maximum shear prior to twinning at T9 is 0.17, while the formation of twins (T10) and cross-twins (T19) increases shear to 0.29 and 0.31, respectively. The reported shear value for twinning in literature [72] for the bcc solids is 0.35.

We also investigated the cyclic loading-unloading from a finite-element method, as implemented in a crystal plasticity module [73]. The elastic constants from MD simulations at 300K from **Table 1** were used as the model input. The stress strain in **Fig. S1(a)** shows similar hysteresis loops under uniaxial tensile loading as in **Fig. 5**, which depicts cyclic softening as the number of cycles increases. With large-scale simulations, we also find twinning and detwinning phenomena under loading and unloading of MWZTT. These findings further suggest that the MWZTT investigated here can have utility as structural material as categorized by Miracle *et al.* [13]. Notably, a marginal drop in stress in Fig. 5a arises from the nucleation of the twinning planes before actual twinning that eventually leads to the first drastic drop in stress encountered at T10. Additionally, in Fig. 5a, a marginal difference in stress values at the end of tensile (T) loading and at the beginning of compressive (C) loading arises from quasi-static loading employed between two successive loadings. A movie showing the nucleation of the twinning planes is added to the supplementary section Fig. S.

A hysteresis like MWZTT in **Fig. 5** was reported recently (both from MD and experiments) in a nanocrystalline tungsten (W) [74]. Wang *et al.* [73] showed that the hysteresis mainly arises from the twinning process. To understand reason for twinning behavior in MWZTT and nanocrystalline W, we investigated the effect of uniaxial loading on bulk-W where we found that the twinning behavior is intrinsically related to bulk-W, see **Fig. 6**. The twinning features in a bulk W were reproducible for [100], [010], and [001] uniaxial loadings. The common neighbor analysis (in the inset of **Fig. 6**) shows the twin region of bulk-W (in white). The elastic-deformation response for elemental W is also isotropic, i.e., Zener anisotropy ratio is 1.

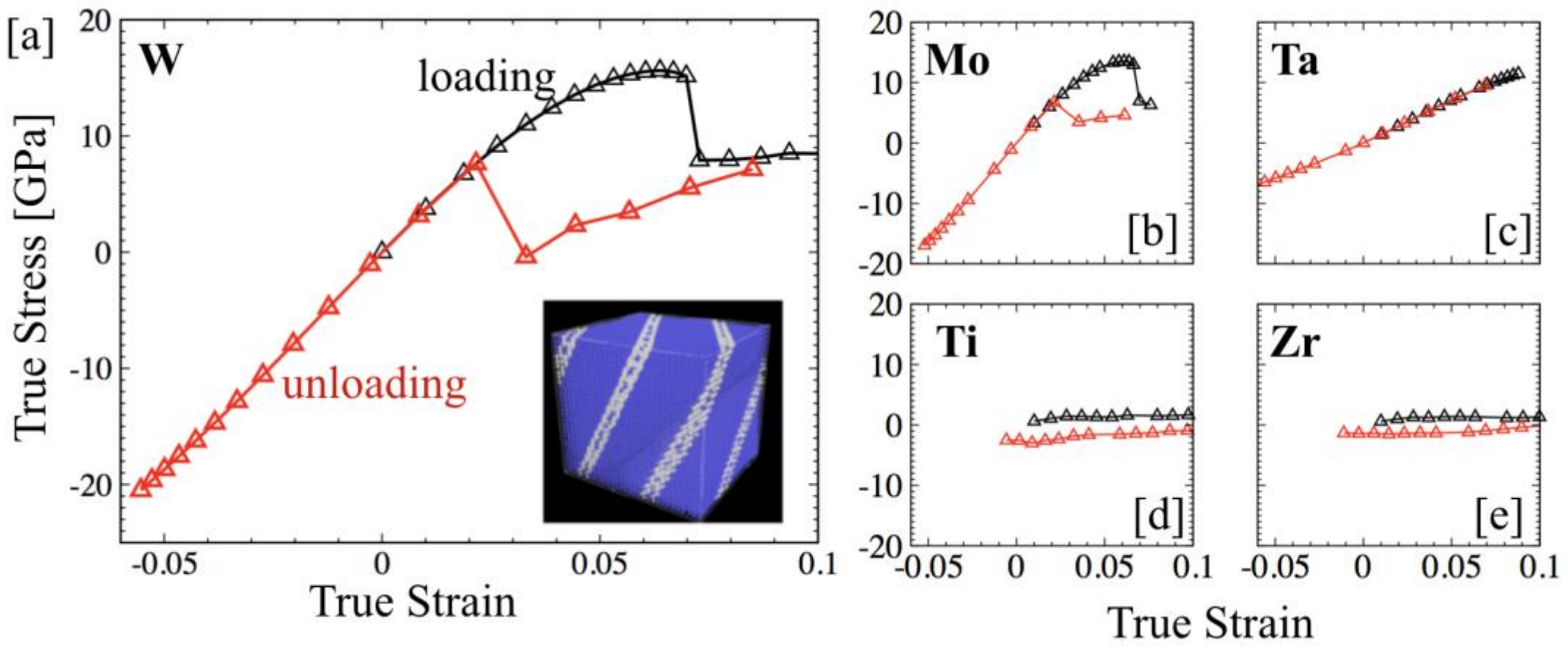

**Figure 6**. (a) Uniaxial loading (tension) and unloading (compression) for W reveals hysteresis in the stress-strain curve at 300 K, which is like MWZTT in **Fig. 5** and nanocrystalline W [74]. [b-e] Stress-strain curve of other constituent elements of MWZTT MPEA, i.e., Mo, Ta, Ti, and Zr.

To understand the critical role of W in MWZTT, we plot stress-strain curve for each uniary alloy under uniaxial quasistatic (un)loading cycle in **Fig. 6b-d.** Except Mo, none of the elements show hysteresis in stress-strain as found for elemental tungsten. The reason for the hysteresis of Mo has similar origin as that of W, as both belong to the same group in the periodic table. This suggests the need of a critical range of W/W-Mo to achieve pseudo-elastic behavior in MWZTT multi-principal element alloys. The stress-strain analysis of equiatomic binary and ternaries in supplemental **Fig. S2 and Fig. S3** also emphasize on the critical role of W/W-Mo to achieve hysteresis in cyclic loading, which is responsible for pseudoelastic behavior. Although Mo in elemental form shows similar hysteresis in stress-strain as W, when mixed in non-W alloys, Mo was not able to retain pseudoelastic behavior.

Previously, pseudoelastic behavior had been observed in a couple of refractory-based systems [75,76]. However, we know of no reports on pseudoelasticity in quinary MPEAs. The deformation mechanisms in the quaternary (Ti-Nb-Ta-Zr) are found to be dependent on composition and vary between slip and twinning modes [75,76]. Increasing the size of the simulation cell up to 50%, we find identical twinning-behavior, as shown in **Fig. 5**. This trend suggests that twinning is inherent feature of MWZTT under uniaxial <100> loading. Formation of twin bands are found during uniaxial loading, while detwinning initiates during unloading. This behavior resembles reports on the bcc tungsten (nanostructured), where the existence of incompatibility at the intersection between the twin bands, and the grain boundary drives detwinning [74]. Another explanation for the driving force for detwinning is based on the instability of the twin structure that increases with the proportion of inclined twin boundaries in the single crystal W [77]. We believe that the deformation modes found in MWZTT are dominated by W and are along the lines of W single crystal. Further experiments are needed to verify the exact driving mechanisms in MWZTT, but as shown here, it would surely relate to the behavior of pure W.

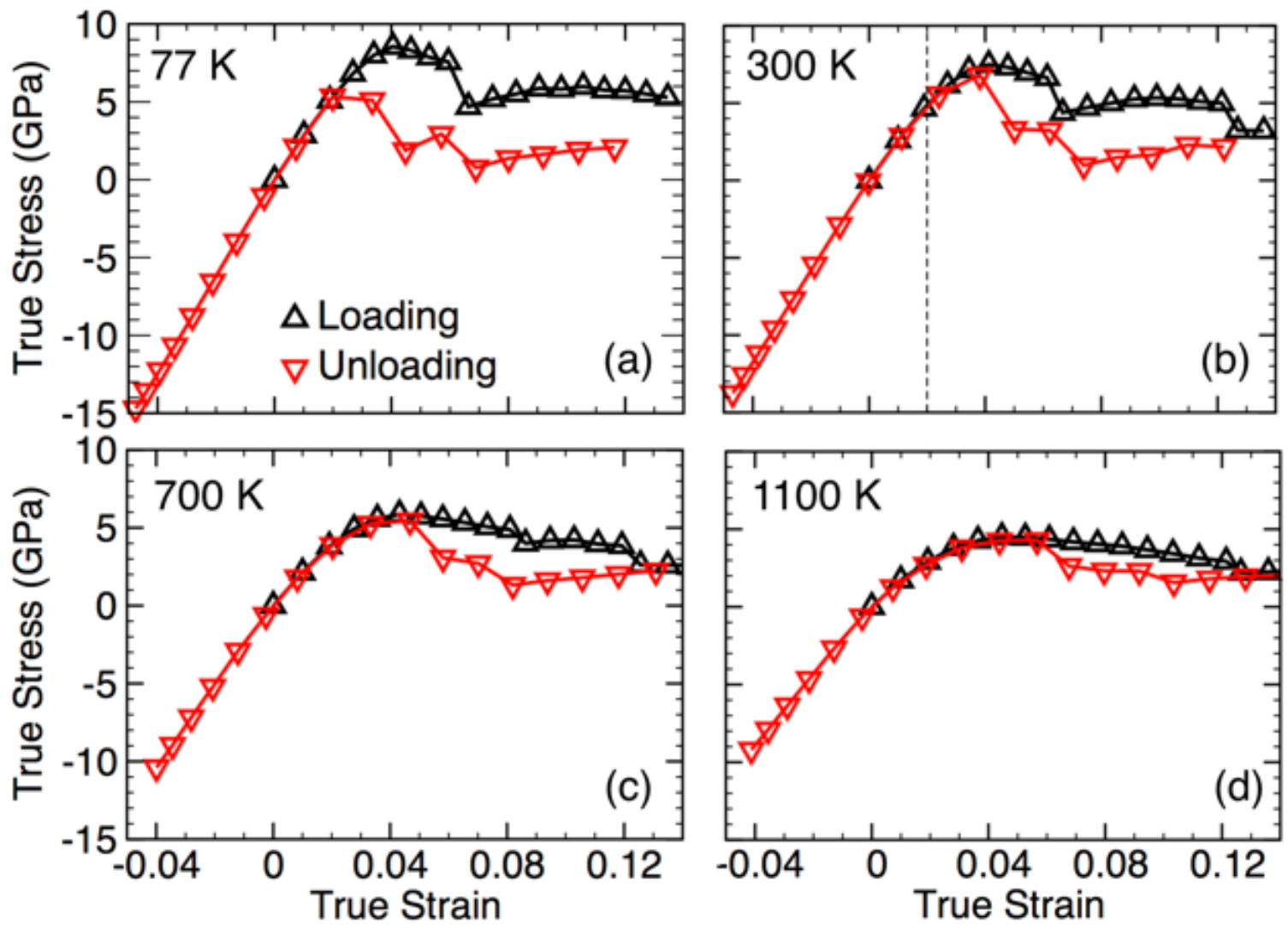

**Figure 7**. Simulated pseudoelasticity for MWZTT vs. temperature: (a) 77, (b) 300, (c) 700, and (d) 1100 K, showing thermal sensitivity. A large hysteresis at 77 K decreases with increasing T until almost vanishing at high temperatures (>1,100 K).

To analyze the effect of thermal fluctuation on the pseudoelasticity in the MWZTT solid-solution, we performed temperature-dependent loading-unloading deformations from 77 K to 1,500 K. The first twin formation is observed near a strain of 0.065, **Figs. 7a-b**, while cross-twins appear near a strain of 0.12. With further increase in temperature, **Figs. 7c-d**, MWZTT allows only cross-twins at a higher strain (0.12), which softens the elastic modes, compared to low-temperature cases [78]. Thus, the large thermal fluctuations at higher temperatures require a greater external strain for twin nucleation. The softening of elastic modes at higher temperatures (**Fig. 7**) leads to a reduction in the loading-unloading hysteresis. In MWZTT, the temperature can be used as a control to tune the pseudoelasticity. Recent studies show that a single-crystal elemental W exhibits deformation twins at low temperatures (T < 273 K, 0$^{o}$C). Such studies also suggest that an increase in the purity would facilitate the twinning process [79]. Our MD analysis indicates twinning during tensile loading for pure W (**Fig. 6**) and MWZTT.

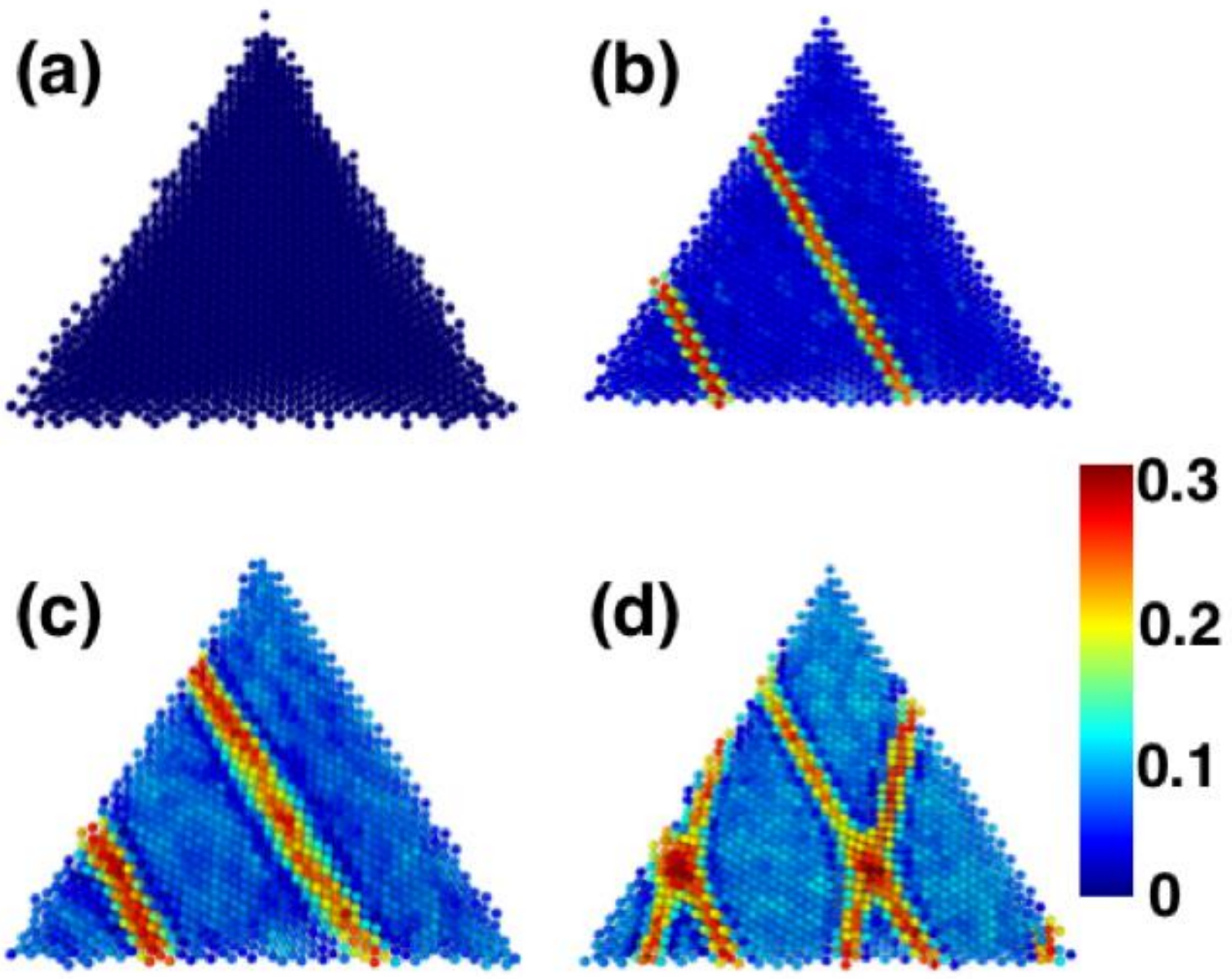

**Figure 8**. Variation of shear strains during loading (tensile) cycle along [111]. (a) T1, (b) T9, (c) T17, and (d) T19 represents quasistatic loading steps at different strains. Twin-layer thickness grows from 3 (T9) to 4 (T17) atomic layers. Further straining yields cross-twinning strain hardening. The magnitude of stress required for initial twin nucleation is higher than that of its propagation.

We also investigated the evolution of twins with shear strain for MWZTT under <100> tensile loading in **Fig. 8**. With no load, as anticipated, atoms experience zero shear (**Fig. 8a**), but, with increase in uniaxial loading (<100>), twins appear (**Fig. 8b**; T9). The twinning layers and the cross-twins are ~3 atomic layers thick, up to ~4 atomic layers at T17 (**Fig. 8c**). Any increase in the quasistatic (<100>) load beyond this point leads to the formation of additional cross-twins (**Fig. 8d**, T19) for MWZTT. Although we have focused on the <100> case, the features of twins, cross-twins, and reverse twinning (detwinning) are reproduced for uniaxial loading along any of the three symmetry directions: [100], [010], or [001]. However, in the finite-strain regime considered here, biaxial loading (i.e., <110>) of the quinary does not show any evidence of twinning and detwinning during the load-unload cycle.

To further emphasize the choice of composition range, we show in stress-strain plot for three high-entropy alloy in **Fig. 9**: (i) equiatomic MoWTaTiZr, (ii) $(MoW)_{80}Zr_{10}(TaTi)_{10}$, and (iii) $(MoW)_{90}Zr_{5}(TaTi)_{5}$ both (later two) along the MWZTT line in **Fig. 2**. The equiatomic alloy does not show any hysteresis, where Mo+W composition is equal to 40at.% in Mo-W-Ta-Ti-Zr MPEA. However, with increase in at.%(Mo+W), we observed some hysteresis character in **Fig. 9b**, however, loading and unloading curve run almost parallel with no reproducibility of closed-loop behavior. In contrast, similar hysteresis was observed in going beyond MWZTT composition as shown in **Fig. 9c**.

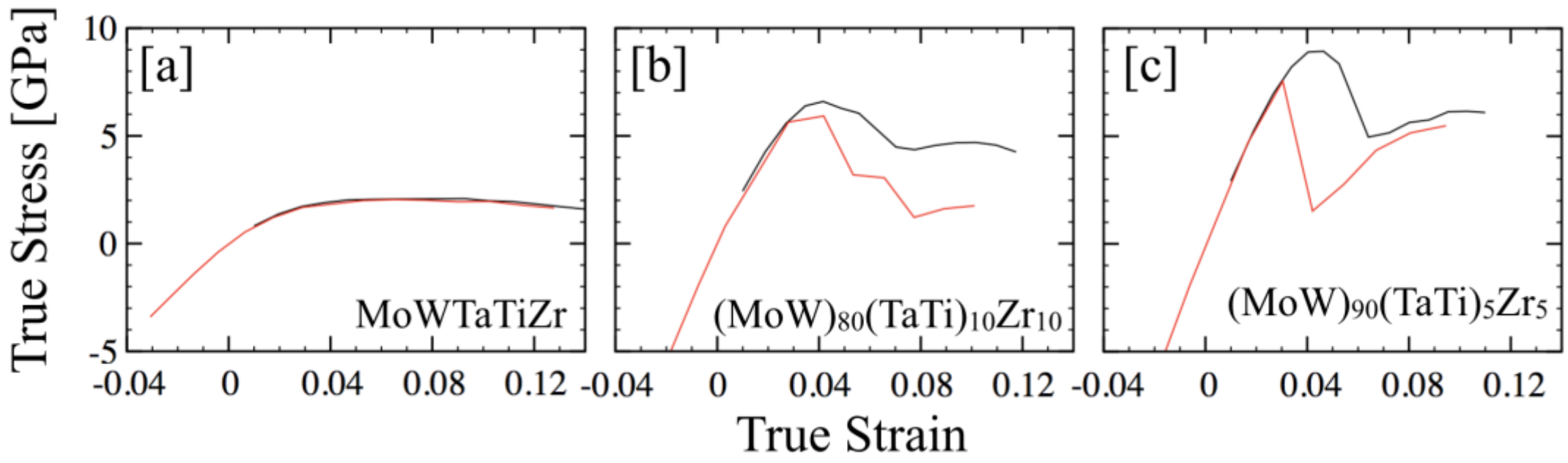

**Figure 9.** Hysteresis under uniaxial {001} quasistatic (un)loading for - (a) MoWTaTiZr, (b) $(MoW)_{80}Zr_{10}(TaTi)_{10}$, and (c) $(MoW)_{90}Zr_{5}(TaTi)_{5}$ at 300 K in Mo-W-Zr-Ta-Ti MPEAs.

Our analysis indicates that at 85at.%(Mo+W) is the critical composition where pseudoelastic behavior is prominent and continues with further increasing at.%(Mo+W). We have also added some critical tests on equiatomic binary and ternary alloys, see **Fig. S2** and **Fig. S3** (also Fig. S4 for phase stability in ternary phase diagram), emphasizing the critical role of W/W-Mo composition range. Individually, Ta, Ti, and/or Zr as such does not display pseudoelastic behavior, however, in Mo-W-rich alloys, they help to stabilize BCC phase, as shown in **Fig. 2**. Importantly, Singh *et al.* [23,27] have shown recently that Ti/Zr and Ta add critical electronic (stabilization) effects in quinary cases. Thus, the choice of specific composition of each constituent element in MWZTT is based on the fact that pseudo-elastic behavior and phase stability is preserved throughout the temperature range.

## Conclusion

We performed detailed phase stability and property analysis of (Mo-W)-Zr-(Ta-Ti) RMPEAs using CALPHAD phase diagrams and DFT-derived formation energies. A critical composition was down selected from the single-phase bcc region in $(MoW)_{x}Zr_{y}(TaTi)_{1-x-y}$, i.e., $(MoW)_{85}$ $Zr_{7.5}(TaTi)_{7.5}$ (denoted MWZTT), satisfying phase-, mechanical-, and structural-stability criteria. We show that only Mo-W-rich compositions, i.e., Mo+W$\geq 85\ at.\%$, show reproducible hysteresis in stress-strain responsible for pseudo-elastic behavior in Mo-W-Ta-Ti-Zr. The MWZTT composition shows robust temperature-dependent elastic properties, and peculiar pseudoelastic behavior under uniaxial loading. The room-temperature finite-element modeling also reproduces similar behavior found in MD simulations, suggesting robustness of the pseudoelasticity in MWZTT. The MD simulations further show a strong temperature dependence of a twinning/detwinning process during loading/unloading (tension-compression) cycle. The detwinning process driven by an instability of the twin structure is the main factor leading to the pseudoelastic behavior, which can be controlled using temperature. Additionally, the finite-element modeling exemplifies the hysteresis loop during cyclic loading and usable yield stress for the proposed RMPEA, see the supplementary material. While this study needs further experimental verification, the proposed refractory-based pseudoelastic material can be designed to operate under specific temperature ranges with specific mechanical behavior, offering new technological opportunities through materials design.

**Acknowledgements**

The work at the Ames Laboratory was supported by the U.S. Department of Energy (DOE), Office of Science, Basic Energy Sciences, Materials Science and Engineering Division. Ames Laboratory is operated for the U.S. DOE by Iowa State University under Contract No. DE-AC02-07CH11358. AS and GB thank the Office of Naval Research (ONR) for support through the grants N00014-16-1-2548 and N00014-18-1-2484. TK acknowledges the support of NSF through Grant No. NSF-DGE-1545403. The MD simulations were supported in part by a grant of computer time from the DoD High-Performance Computing Modernization Program at Army Engineer Research and Development Center. Phonon calculations and MD simulations in part were carried out at the Texas A&M High-Performance Research Computing (HPRC) Facility.

**CRediT authorship contribution statement**

**Aayush Sharma**: Conceptualization, MD Simulations, Formal analysis, Writing – original draft, Writing – review & editing. **Prashant Singh**: Conceptualization, DFT calculations, MD Simulations, Formal analysis, Supervision, Writing – original draft, Writing – review & editing. **Tanner Kirk**: CALPHAD calculations, Writing – CALPHAD part, Writing – review & editing. **Vellery I. Levitas**: Writing – review & editing. **Peter K. Liaw**: Writing – review & editing. **Ganesh Balasubramanian**: Funding acquisition, Resources, Writing– review & editing. **Raymundo Arroyave**: Funding acquisition, Resources, Writing– review & editing. **Duane D. Johnson**: Funding acquisition, Supervision, Resources, Writing– review & editing.

**Data availability**

The authors declare that the data supporting the findings of this study are available within the paper and supplement. Also, the data that support the plots within this paper and other finding of this study are available from the corresponding author upon reasonable request.

## Declaration of Competing Interest

The authors declare that they have no known competing financial interests or personal relationships that could have appeared to influence the work reported in this paper.

## References.

Supplemental Materials

# Pseudoelastic deformation in Mo-based refractory multi-principal element alloys

A. Sharma,[1,2,*] P. Singh,[1,*] T. Kirk,[3] V.I. Levitas,[4,5,6] P.K. Liaw,[7] G. Balasubramanian,[8] R. Arroyave,[3,9] and Duane D. Johnson[1,6]

[1]*Ames Laboratory, U.S. Department of Energy, Iowa State University, Ames, IA 50011 USA*
[2]*Sandvik Coromant R&D, Stockholm, 12679 Sweden*
[3]*Department of Materials Science & Engineering, Texas A&M University, College Station, TX 77843, USA*
[4]*Department of Mechanical Engineering, Iowa State University, Ames, IA 50011 USA*
[5]*Department of Aerospace Engineering, Iowa State University, Ames, IA 50011 USA*
[6]*Department of Materials Science & Engineering, Iowa State University, Ames, IA 50011 USA*
[7]*Department of Materials Science & Engineering, The University of Tennessee, Knoxville, TN, 37996 USA*
[8]*Department of Mechanical Engineering & Mechanics, Lehigh University, Bethlehem, PA 18015 USA*
[9]*Department of Mechanical Engineering, Texas A&M University, College Station, TX 77843, USA*

**KKR-CPA Method details**: The KKR-CPA is an effective-medium theory that solves for an effective-scatterer at a site such that we recover the scattering properties on average for the collective disordered system, i.e., the KKR-CPA is an effective scatterer theory with energy-dependent complex values (Real & Imaginary). As a result, the effects of disorder are represented by broadening the electronic structure (not possible in an effective-atom method), leading to widths in k- and E-, related to electron-scattering loss into other scattering channels. The width in **k**, for example, is directly related the electron-scattering length and permits quantitative calculation of, e.g., resistivity changes. Without disorder, the electron dispersion would have zero-width peaks (delta functions), as in ordered alloys. Disorder varies with energy and the dispersion width varies as such. Importantly, the Green's function is directly calculated giving correct spectral average properties that directly compare to experimental features. These features are all discussed in referenced work and pointed out in the manuscript [48].

**Phonon and elastic property calculation**: We designed an 80-atom disorder supercell [50] to mimic $(MoW)_{85}Zr_{7.5}(TaTi)_{7.5}$ using 2-atom bcc unit cell periodically in a 5 x 4 x 2 lattice to calculate elastic parameters (Cij) and phonons using first-principles density functional theory.

**Supplementary Table 1**. The lattice constants of fully relaxed MWZTT disorder supercell are a = 15.90946, b = 12.774787, and c = 6.39163. The atomic co-ordinates for Mo (n = 34), W(n = 34), Zr(n = 6), Ti(n = 3), and Ta(n = 3) are also given.

| Atom Type | X | Y | Z |
|---|---|---|---|
| Mo1 | 0.80224451 | 0.00019038 | 0.99568459 |
| Mo2 | 0.99872478 | 0.25125761 | 0.99935253 |
| Mo3 | 0.99928191 | 0.50090635 | 0.99963125 |
| Mo4 | 0.60078367 | 0.50063422 | 0.0001441 |
| Mo5 | 0.79999007 | 0.50043186 | 0.99957552 |
| Mo6 | 0.00029961 | 0.74849513 | 0.99963863 |
| Mo7 | 0.19910686 | 0.99835683 | 0.50047133 |
| Mo8 | 0.39855227 | 0.99845068 | 0.50090516 |
| Mo9 | 0.80281559 | 0.00046487 | 0.50462106 |
| Mo10 | 0.19561992 | 0.24539085 | 0.5000781 |
| Mo11 | 0.40245502 | 0.24620911 | 0.50013837 |
| Mo12 | 0.60081559 | 0.25383844 | 0.49994809 |
| Mo13 | 0.40344639 | 0.50538198 | 0.50048079 |
| Mo14 | 0.80059459 | 0.50023896 | 0.49992409 |
| Mo15 | 0.19988128 | 0.75027792 | 0.50098068 |
| Mo16 | 0.59886243 | 0.74611092 | 0.5026032 |
| Mo17 | 0.09785001 | 0.12364082 | 0.2544919 |
| Mo18 | 0.30145635 | 0.12045113 | 0.25409977 |
| Mo19 | 0.7023282 | 0.13058243 | 0.25176563 |
| Mo20 | 0.09688834 | 0.37803925 | 0.25208012 |
| Mo21 | 0.50199225 | 0.37726293 | 0.24998665 |
| Mo22 | 0.09949141 | 0.62559179 | 0.24960553 |
| Mo23 | 0.29972472 | 0.62716243 | 0.25067209 |
| Mo24 | 0.70023899 | 0.62247126 | 0.24965327 |
| Mo25 | 0.0981232 | 0.12385592 | 0.74353395 |
| Mo26 | 0.30148225 | 0.12090853 | 0.7453357 |
| Mo27 | 0.70168887 | 0.1294955 | 0.74788475 |
| Mo28 | 0.9001291 | 0.12423637 | 0.74759253 |
| Mo29 | 0.50187229 | 0.37720721 | 0.75009754 |
| Mo30 | 0.70002062 | 0.37776599 | 0.74858171 |
| Mo31 | 0.09953104 | 0.62647452 | 0.7491965 |
| Mo32 | 0.30018402 | 0.6268851 | 0.75038726 |
| Mo33 | 0.70019712 | 0.62311205 | 0.74978856 |
| Mo34 | 0.49706159 | 0.8702841 | 0.74992137 |
| W1 | 0.39812903 | 0.99843663 | 0.99916912 |
| W2 | 0.40369005 | 0.24686498 | 0.00026135 |
| W3 | 0.60061904 | 0.25383197 | 0.999851 |
| W4 | 0.19658643 | 0.5071068 | 0.99934172 |
| W5 | 0.40345504 | 0.50519844 | 0.0008035 |
| W6 | 0.19976943 | 0.75067244 | 0.99911201 |
| W7 | 0.39957333 | 0.75031576 | 0.00012378 |
| W8 | 0.59930392 | 0.74615431 | 0.99689486 |
| W9 | 0.80233603 | 0.74592308 | 0.99704483 |
| W10 | 0.0013509 | 0.99893344 | 0.50069589 |
| W11 | 0.99941094 | 0.25129766 | 0.50126362 |
| W12 | 0.79959232 | 0.25169186 | 0.50125038 |
| W13 | 0.99909064 | 0.5003007 | 0.49953692 |
| W14 | 0.19549515 | 0.50586223 | 0.49985936 |
| W15 | 0.60137095 | 0.49972356 | 0.50043374 |
| W16 | 0.00039997 | 0.7494998 | 0.49964534 |
| W17 | 0.80314293 | 0.74642063 | 0.50341279 |
| W18 | 0.49697856 | 0.13066167 | 0.25003691 |
| W19 | 0.69985792 | 0.37811643 | 0.25161556 |
| W20 | 0.89976465 | 0.37670695 | 0.25103968 |
| W21 | 0.50079702 | 0.62341994 | 0.25010912 |
| W22 | 0.90059119 | 0.62373388 | 0.24955758 |
| W23 | 0.1012963 | 0.87193542 | 0.25144696 |
| W24 | 0.49537226 | 0.86986989 | 0.24979435 |
| W25 | 0.90117976 | 0.87273311 | 0.25075823 |
| W26 | 0.49657732 | 0.13062652 | 0.74959775 |
| W27 | 0.09663448 | 0.37820904 | 0.74657724 |
| W28 | 0.90011424 | 0.37694565 | 0.74801876 |
| W29 | 0.5011881 | 0.62364635 | 0.75245805 |
| W30 | 0.90044585 | 0.62458128 | 0.74985697 |
| W31 | 0.1003528 | 0.87221008 | 0.74661584 |
| W32 | 0.29832192 | 0.8743668 | 0.75125355 |
| W33 | 0.70513085 | 0.86901966 | 0.7505796 |
| W34 | 0.90110091 | 0.87246047 | 0.74865349 |
| Zr1 | 0.59844858 | 0.00130394 | 0.99660485 |
| Zr2 | 0.19743407 | 0.2447695 | 0.0005247 |
| Zr3 | 0.59913301 | 0.00207416 | 0.5030653 |
| Zr4 | 0.30075439 | 0.37673347 | 0.25432724 |
| Zr5 | 0.70284128 | 0.8715916 | 0.25147968 |
| Zr6 | 0.30039912 | 0.37666411 | 0.74593884 |
| Ti1 | 0.19942811 | 0.99857241 | 0.99832625 |
| Ti2 | 0.39964422 | 0.74959792 | 0.50412856 |
| Ti3 | 0.29843817 | 0.8744574 | 0.25268384 |
| Ta1 | 0.00214591 | 0.99822507 | 0.99786642 |
| Ta2 | 0.79847928 | 0.25160174 | 0.99697656 |
| Ta3 | 0.90009879 | 0.12486783 | 0.25255561 |

*Finite Element Modeling (FEM):* Cyclic loading within the PRISMS Crystal plasticity module [75] was used to analyze the stress-strain response for the MWZTT MPEA. Here, the elastic constants at 300K from the MD simulations was used to calibrate the model input parameters. The relative linear solver tolerance was set as 1.0 x $10^{-10}$, and the maximum non-linear iterations was limited to 4, with a time-increment as 0.1 and a total simulation time of 100. The number of slip systems was 12, with the initial slip resistance of 200 MPa, initial hardening modulus as 1,500 MPa, and power law coefficient as 1, the saturation stress was 500 MPa. The stress tolerance for the yield surface was 1.0 x $10^{-9}$, and maximum slip search iteration and maximum number of iterations to achieve non-linear convergence are both restricted to 1. The input microstructure had 20 x 20 x 22 voxels in each of the 3 directions (x, y, and z). For the displacement, the face-id of 2 with 1 degree of freedom and 0.1 as the final displacement was chosen.

The stress-strain plot in **Fig. S1a** shows cyclic stress-strain hysteresis loops for the MWZTT alloy, the yield stress obtained from **Fig. S1b** is 500 MPa. These findings further suggest that the MWZTT alloy investigated in the present work surely has utility as a structural material. The possibility of tunable pseudoelastic properties is very intriguing, and it could be needed to be verified through experiments. The experiments and further detailed FEM study are not in the scope of the present work and would be presented elsewhere.

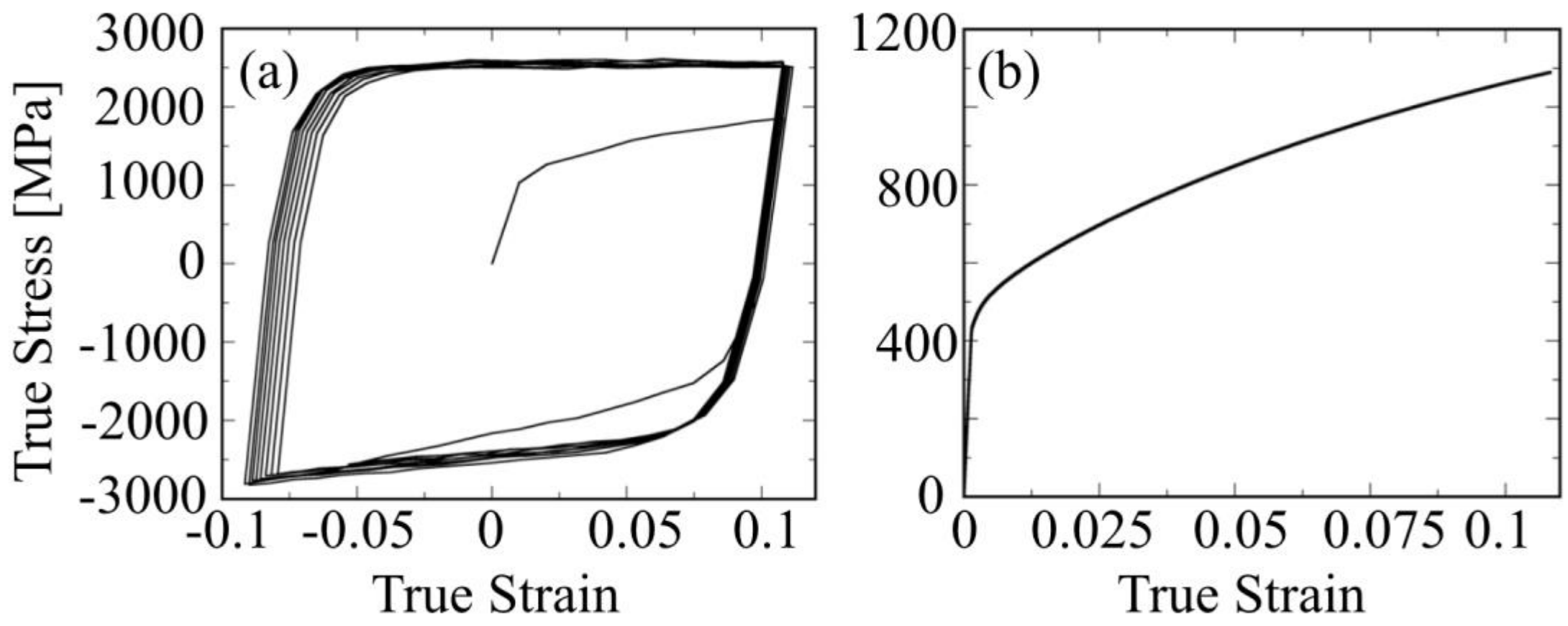

**Figure S1.** (a) Cyclic loading characteristics, and (b) tensile loading of the MWZTT MPEA. Crystal plasticity implementation within PRISMS package was used to model the stress-strain response [75].

The marginal drop in stress arises from the nucleation of the twinning planes before actual twinning as shown in **supplemental Movie S1** that eventually leads to the first drastic drop in stress encountered in Fig. 5a

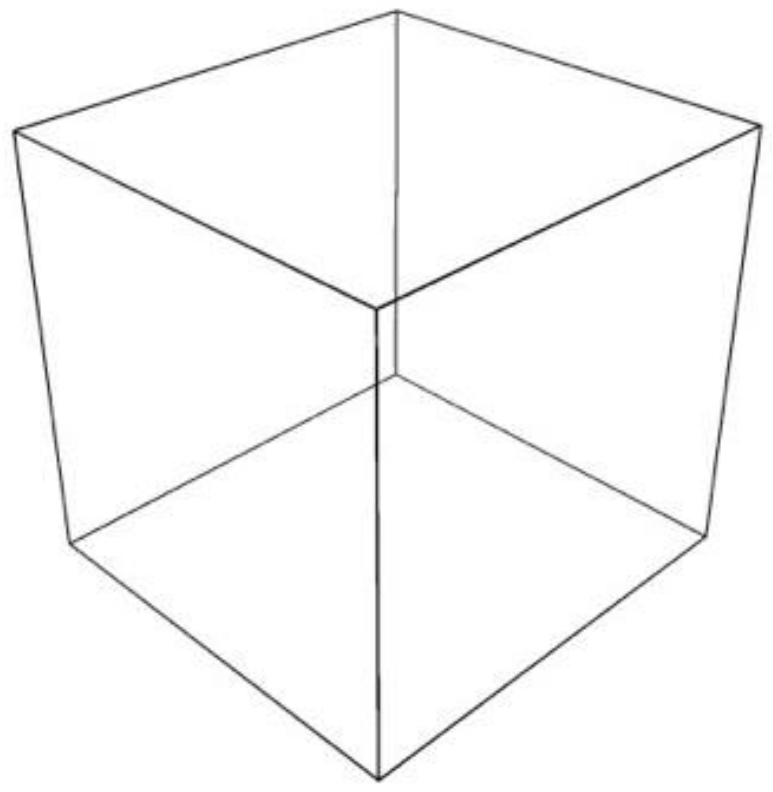

**Supplemental Movie S1:** Movie showing the nucleation of the twinning planes in MWZTT MPEA.

**<u>Pseudo-elastic behavior in binary, and ternary alloys</u>**: To emphasize the critical role of W in enhancing pseudoelastic behavior, as well as detrimental effect of excess Ti/Zr/Ta, we plot hysteresis under uniaxial quasistatic (un)loading cycle for binary, and ternary alloys in **Fig. S2 and Fig. S3**, respectively. None of the elements were found to follow hysteresis of elemental W under uniaxial loading with the exception of Mo. The reason for hysteresis of Mo has similar origin to that of W, as both belong to same group of the periodic table. Despite Mo showing some hysteresis effect, we can clearly see the difference in loop with W at 300 K. This suggests towards need of critical amount of W to achieve pseudo-elastic behavior in MWZTT MPEAs.

To further emphasize the role of W in playing critical role in enhancing pseudo-elastic behavior as well as detrimental effect of excess Ti/Zr/Ta, we plot hysteresis under quasistatic (un)loading cycle for binary WMo, WTi, WZr, and WTa at 300 K in **Fig. S2**. Clearly, excess Ti/Zr disturbs the pseudo-elastic nature of W-based alloys. Although Ta maintains the shape of the hysteresis, but it impacts the close loop behavior found in W rich alloys.

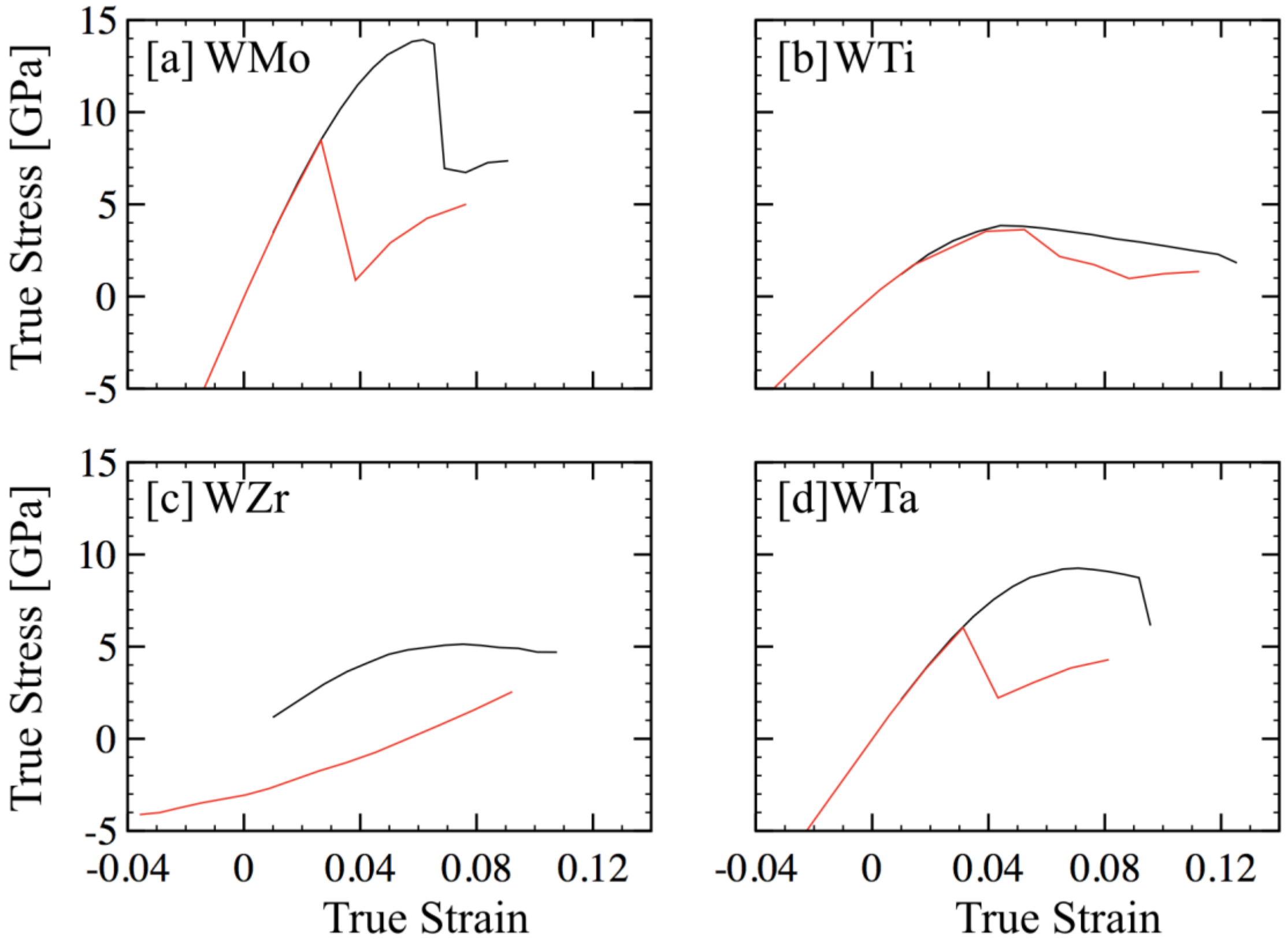

**Figure S2.** Hysteresis under uniaxial {001} quasistatic (un)loading cycle of binary (a) WMo, (b) WTi, (c) WZr, and (d) WTa at 300 K.

To add more clarity on the effect of W, we performed MD simulations on equiatomic without-W (TaTiZr, MoTaTi) and with-W (WMoTa, WTaTi, and WTiZr) alloys, and plot stress-strain behavior in **Fig. S3**. The alloys with without-W show no hysteresis in stress-strain curve, i.e., cyclic loading behavior is absent. Although, elemental Mo shows hysteresis in **Fig. 6b,** but it could not maintain the same behavior when mixed with other alloying elements in absence W.

On the other hand, the alloys with-W in **Fig. S3**, i.e., WMoTa and WTaTi show clear hysteresis, however, WTiZr shows broken hysteresis. This suggests towards detrimental role played by excess Ti+Zr, which further confirms the need of critical at.%(Mo+W) or at.%W in order to realize the pseudo-elastic behavior. All these calculations indicate towards one common point that W/W-Mo is a necessary component to achieve pseudoelasticity in Mo-W-Ta-Ti-Zr MPEA.

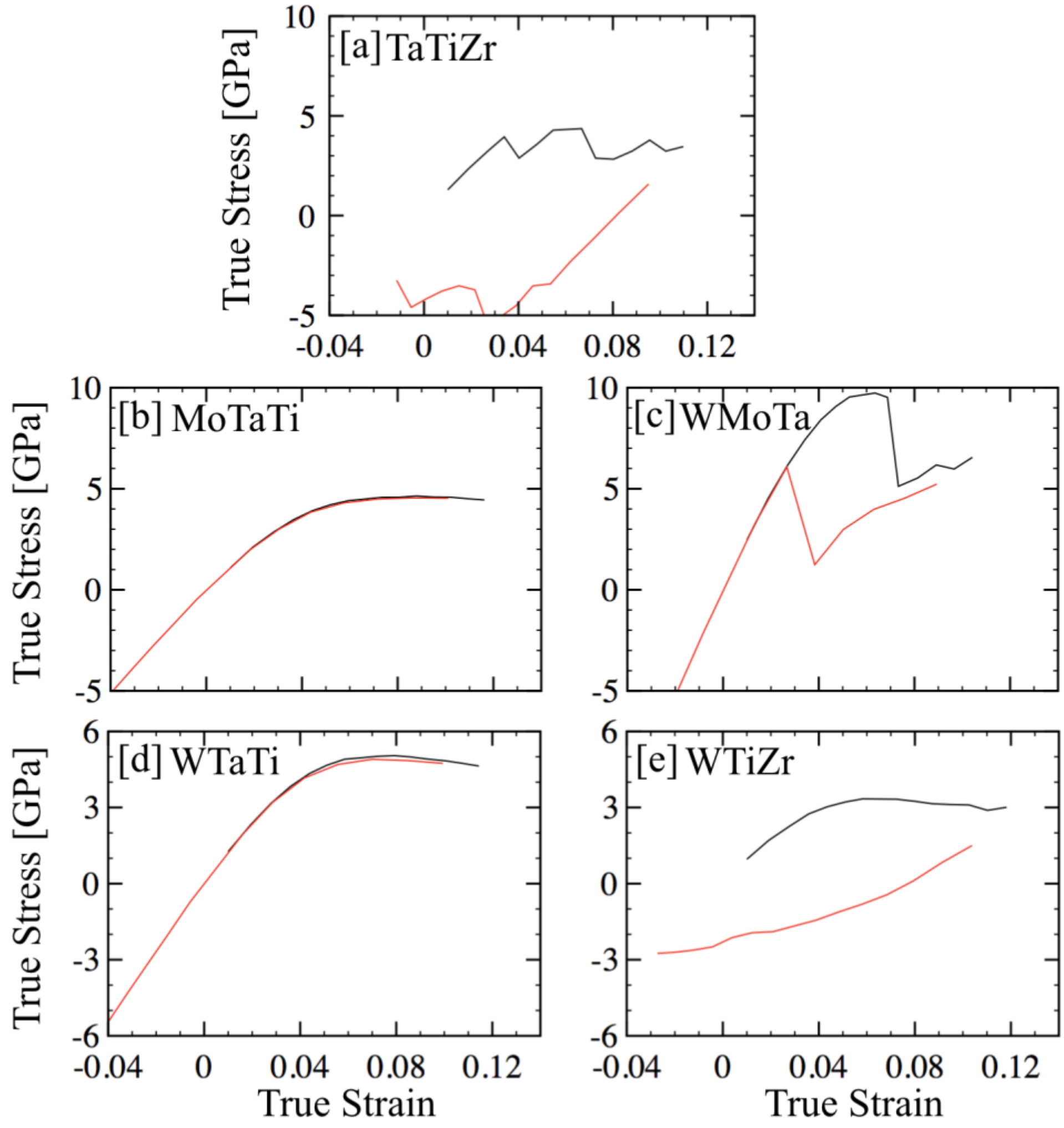

**Figure S3.** Hysteresis under uniaxial {001} quasistatic (un)loading cycle for equiatomic (a) TaTiZr, (b) MoTaTi, (c) WMoTa, (d) WTaTi, and (e) WTiZr at 300 K.

In **Fig. S4,** we show the CALPHAD phase stability of W/W-Mo/W-Ta/Mo-Ta based ternary alloys from Mo-W-Ta-Ti-Zr family as in **Fig. S3**. Stability analysis shows that it is only W-Mo-Ta family alloy, i.e., equiatomic WMoTa, that shows hysteresis in stress-strain curve in **Fig. S3c**. Clearly, W+Mo range of 67at.% in this family of alloys further affirms our proposed minimum range of W or W+Mo based alloys in order to achieve hysteresis in stress-strain, i.e., psuedoelastic behavior. For example, Mo that shows hysteresis in stress-strain in elemental form, loses this character when mixed with other constituent elements. Only W-rich or W+Mo-rich alloys were found with pseudo-elastic character. We note that the contributions of the other elements (Ta, etc.) is to enhance the phase stability of the BCC phase due to entropic effects.

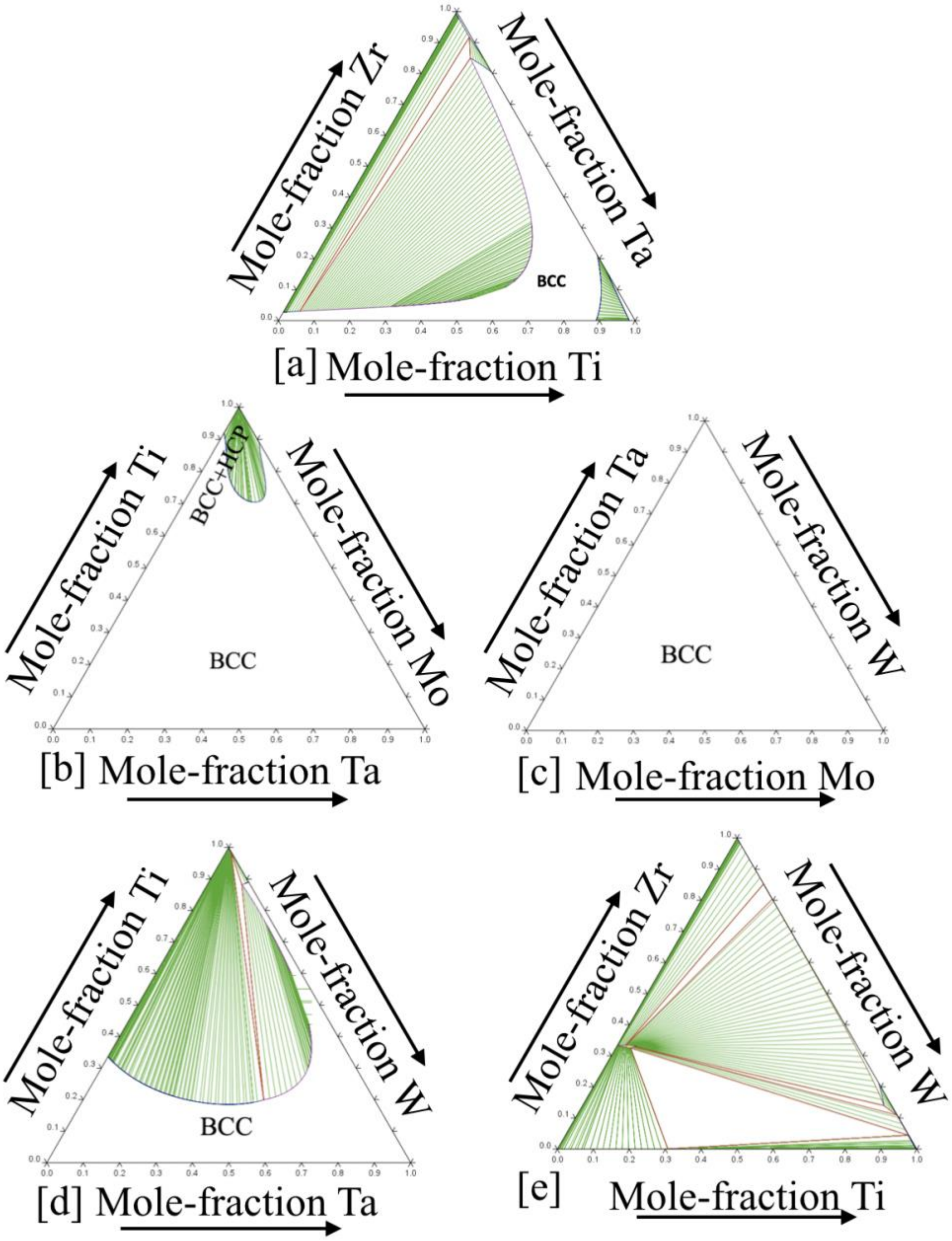

**Figure S4**. High-temperature (1000 K) CALPHAD phase-diagram analysis of (a) TaTiZr, (b) MoTaTi, (c) WMoTa, (d) WTaTi, and (e) WTiZr.